%% file: Tunneling.tex
\documentclass{article}
\usepackage[dvips]{graphicx}
\usepackage{amsmath,amsbsy,amsfonts,amssymb}
\input{Definitions.tex}

\begin{document}

\begin{center}
\Large{TUNNELING EFFECTS BETWEEN TORI IN DOUBLE WELLS}%
\footnote{to be presented as a poster in the session on Quantum Mechanics and Spectral Theory at ICMP 2003.}
\end{center}
\bigskip
\begin{center}
Michel ROULEUX
\end{center}

\begin{center}
Centre de Physique Th\'eorique

Unit\'e Propre de Recherche 7061

CNRS Luminy, Case 907, 13288 Marseille Cedex 9, France

and PhyMat, Universit\'e de Toulon et du Var
\end{center}


\bigskip
\noindent {\bf Abstract}: We consider tunneling between 2 symmetric potential 
wells for a 2-d Schrodinger operator, in the case of eigenvalues 
associated with with quasi=modes 
supported on KAM or Birkhoff tori.
\medskip

\noindent {\bf 0. Introduction}. 
\smallskip
We consider
here tunneling between 2 symmetric potential wells for a 2-d
Schr\"odinger operator $P=-h^2\Delta+V$ in the limit $h\to 0$,
near some energy level $E_0=0$ close to the non degenerate minima of $V$.

Tunneling is a difficult problem that has exercised so far many subtle
and ingenious strategies ; at least, computing tunneling rates 
involves various
scenarios which depend on the details of the dynamics, ranging from
integrable or quasi-integrable systems, to ergodic or chaotic ones
(see [W], and [Cr] for a recent review. )

As a general rule, the energy shift (or splitting of eigenvalues)
is related to the so called Agmon distance $S_0(E)$ between the wells,
associated with the degenerate, conformal metric $ds^2=(V-E)_+dx^2$
that measures the life-span (instanton)
of the particle in the classically forbidden region $V(x)\geq E$. 

Much is known in the 1-d case, even for excited states, or in several
dimensions for the lowest eigenvalues. For general wells, there is 
the following equivalence [Ma]~:  the corresponding 
normalized eigenfunctions
are non exponentially small 
(i.e. for all $\e >0$, larger, in local $L^2$ norm, than a constant times
$e^{-\e /h}$, 
$0<h\leq h_{\e }$, ) 
where minimal geodesics, connecting the 2 wells, 
meet their boundary, if and only if the splitting
is non exponentially small with respect to $e^{-S_0(E)/h}$.

We study the special case of 
splitting of eigenvalues associated with quasi-modes supported
on KAM or Birkhoff tori~; our goal is to compute tunneling rates
for a large family of such eigenvalues, which we shall call
a {\it spectral tunnel series}. 
\medskip

\noindent {\bf 1. Tori and quasi-modes}.
\smallskip
Let us consider for a moment the case of a single well around $U_0=0$ 
(so that we can ignore interaction with the other well, ) and
let $p_0(x, \xi)=\xi^2+\lambda_1^2x_1^2+\lambda_2^2x_2^2$ be the 
quadratic part of the (smooth) classical hamiltonian $p(x, \xi)$ near 0
(quadratic approximation). If the frequencies $\lambda_j$'s are rationnally 
independent, then Birkhoff's theorem 
tells us that the orbits of $p$ near the fixed point $(x, \xi)=0$
with energy $E$ are quasi-periodic, in the sense that they are
confined within quasi-invariant tori (the Birkhoff tori,) over a time scale
${\cal O}(E^{-\infty})$. 
Whenever the system is non integrable, most of these tori, 
will be destroyed and replaced by chaotic regions~; however  
(under a suitable a diophantine condition 
on the frequencies, ) the so-called KAM tori,
whose collection form a   
Cantor set, eventually survive.  

Such lagrangian (or possibly isotropic)
integral manifolds $\Lambda_\iota^E$ support quasi-modes, whose energies
are given by the Einstein-Brillouin-Keller (EBK) quantization
rule~: if $H(\iota)=2\lambda_1\iota_1+2\lambda_2\iota_2+\cdots$ 
is the hamiltonian 
expressed (asymptotically) in terms of action-angle variables 
$(\iota, \varphi)$, then the energies of the eigenstates 
are given to first order in $h$ by~:
$$E_\alpha(h)=H(\iota_\alpha), \quad \iota_\alpha=(\alpha+{1\over 4}\nu)h,
\quad \alpha\in{\bf N}^2, \ |\alpha|h \leq E_0 \leqno (1)$$
The vector $\nu=(2,\cdots, 2)$ is Maslov index, counting the number
of caustics met along an orbit, passing from sheet to sheet on the torus.

For Birkhoff tori associated with energies $Ch\leq E\leq h^\delta$, 
where $0<\delta<1$, 
EBK formula (1) can be corrected to all orders in $h$
with an accuracy ${\cal O}(h^\infty)$ [Sj]. In case of KAM tori, 
and larger energies (in an interval independent of $h$, )
these expansions hold modulo ${\cal O}(e^{-1/Ch^{1/s}})$, for some $s>1$
related to the diophantine condition on the ratio $\lambda_1/\lambda_2$ [Po]. 
Of course in that case, analyticity properties of the potential
are required as in usual KAM procedures,
which makes the analysis more subtle. But the geometry is the same.

We have seen that tunneling rates hinge upon
decay of quasi-modes at the edge of the well $\partial U_E=
\{V(x)=E\}$. They also 
depend 
on the decay near the caustics of
$\Lambda_\iota^E$, $\iota=\iota_\alpha$. 

The caustics can be viewed as a rectangle shaped fold line 
delimiting the zone of pure
oscillations of the quasi-modes, and touching the boundary of the wells
at 4 vertices, the hyperbolic umbilic points (HU) points,
section of the torus by the plane $\xi=0$ in ${\bf R}^4$. 
According to (1), the edges of this rectangle have a size
$\iota_\alpha=(\alpha+{1\over 4}\nu)h$. 

All tori $\Lambda_\iota^E$ continue analytically in the $\xi$ variables.
Over the classically
forbidden region, analytic continuation amounts
to parametrize the orbits with imaginary time. It is convenient 
to view $\Lambda_\iota^E$ as a 
multidimensional Riemann sheet structure, with a number of sheets 
corresponding to the choice of the sign of momentum, 
gluing along the caustics, and all intersecting
at the HU's. $\Lambda_\iota^E$ is parametrized by a 
phase function $F_y^E(x)$, where $y\in\partial U_E$ denotes the umbilic
and can be identified with $\iota$. 
This phase is complex 
(reflecting the oscillations of the corresponding quasi-mode)
on all but one sheet of $\Lambda_\iota^E$, denoted by $\widetilde
\Lambda_\iota^E$, which lies over
the classically forbidden region, and corresponds to a pure 
exponential decay of the quasi-mode (we exclude the sheet which gives
exponential growth.)

In case of Birkhoff tori, we gave a complete asymptotic 
expansion of the quasi-mode in some region of 
the decaying zone [KaRo], close to, but at a finite distance from the caustics.
Similar expansions can be obtained closer to the caustics,
in term of special functions of Airy type. Yet another expansion
could be found in case of KAM tori.

\medskip
\noindent {\bf 2. Continuation of action integrals in the classically
forbidden region}.
\smallskip
Another central geometric figure of the problem is the integral
manifold of $q(x, \xi)=-p(x, i\xi)$ passing above $\partial U_E$, i.e.
$$
\Lambda_\partial^E = \{ \exp t H_q (\rho) : \rho \in
\partial U_E \times 0, \ q(\rho) = -E, \ t \in {\bf R} \}
$$
This is (locally) a smooth real lagrangian submanifold, of the 
form $\xi = \nabla d_E(x)$, $x\notin U_E$. 
where $d_E(x) = d_E(x, \partial U_E)$ is Agmon
distance from $x$ to $\partial U_E$. 
Actually, 
$\Lambda_\partial^E$ has the fibre bundle structure
$\Lambda_\partial^E =
\bigcup_{y \in \partial U_E}
\gamma_y $.
Here $\gamma_y$ is the bicharacteristic of $q(x, \xi)$ at energy $-E$
issued from $\partial U_E$ at the umbilic $y$, and 
$\gamma_y = \widetilde \Lambda_\iota^E \cap \Lambda_\partial^E$.

Introducing appropriate coordinate charts of 
hyperbolic action-angle variables $(\iota', \varphi)\mapsto (\iota, \varphi')$
given by Birkhoff 
transformations,
we can also view $F_y^E(x)$ as the action $\int_y^x \xi dx$ computed
along some path keeping $\iota'$ constant, and varying $\varphi$.
\smallskip
$\bullet$ Our first task is to continue $x\mapsto F_y^E(x)$ from 
$\widetilde \Lambda_\iota^E$, keeping $y$ and $E$ fixed.

First we introduce some scaling factors. One difficulty throughout
consists in the range of different scales. 
So let $\mu=\sqrt E$ be the characteristic size
of the (euclidean) diameter of $U_E$.  

Let $y=(y_1, y_2)$ be an umbilic, and assume that the torus
$\Lambda_\iota^E$ is not ``too flat'' in a certain sense, or
equivalently, that the rectangle shaped caustics is ``not too far
from being a square''. For $x$ also close enough from $\gamma_y$, so that
$F_y^E(x)$ is still real, we denote by $y(x)$ the unique point of 
$\partial U_E$ such that $x\in \gamma_{y(x)}$.
We have~:
\smallskip
\noindent {\bf Proposition 1}: $F_y^E(x)$ equals $d_E(x)$ precisely
along the geodesic $\gamma_y$. Moreover 
$$d_E(x)-F_y^E(x)\sim {\partial ^2\over \partial y_1^2}F_y^E(x)|_
{x\in \gamma_y} \bigl(y_1-y_1(x)\bigr)^2 
\sim -K(x, y)\bigl(y_1-y_1(x)\bigr)^2$$
where $K(x, y)\geq K_0>0$ whenever  $\dist (x, \partial U_E)\leq C \mu^{1/2}$.
(here dist stands for the euclidean distance.)
\medskip
Denote by $\Gamma_y(x)$ the orthogonal projection of $x$ onto $\gamma_y$.
Using a variant of Gauss Lemma (the geodesic flow is locally a
radial isometry), Proposition 1 shows that there are smooth 
level surfaces 
$N_\mu(s)=\{d_E(x)=s \mu \}$ , $s_1\leq s\leq s_2$, 
$\dist \bigl(N_\mu(s), \partial U_E) \sim \mu^{1/2}$, such that 
$$d_E(x)-F_y^E(x)\sim -{1\over \mu} \bigl(x-\Gamma_y(x)\bigr)^2,
\quad x\in N_\mu(s)\leqno (2)$$ 
Using that eikonal equation is satisfied 
by both $d_E(x)$ and $F_y^E(x)$,
estimate (2) continues in the large, all along $\gamma_y$, so far as
$\gamma_y$
does not reach any caustics. 
This holds in particular, if 
$\gamma=\Upsilon_E$ is a minimal $d_E$-geodesic between $N_\mu(s)$
and a fixed $x_0$, somewhere in between the 2 wells.
\smallskip
$\bullet$ Our second task is to compute action from continuation of energy
surfaces, i.e. by varying $E$ (and $\iota'$ accordingly) but keeping
$x$ fixed. Let $\bigl(x, \xi(x)\bigr)=\bigl(x, \nabla_xd_E(x)\bigr))\in 
\Lambda_\partial^E$,
and $(z', \zeta')=\kappa \bigl(x, \xi(x)\bigr)$, where $\kappa$ 
is a suitable canonical transform related to the mapping
$(\iota', \varphi)\mapsto (\iota, \varphi')$, that preserves the
boundary of the well~:
$\kappa (y, 0)=(y', 0)$.
  
In fact, we shall compute $d_E(x)$ in a ($\mu$-independent) neighborhood 
$\omega$ of 
$\partial U_E$
from Agmon distance $d_0(x)$ at energy 0,
which is known to be a $C^\infty$ function of $x$. We have~:
\smallskip
\noindent {\bf Proposition 2}: For $x\in \omega$,
$d_E(x)=d_0(x)+\Sum_j \iota'_j
\log {z'_j\over y'_j}+{\cal O}(\mu^2)$.
\medskip
Typically, for $x\in \omega$, $\iota'_j\log {z'_j\over y'_j}$ 
is comparable to 
$\mu^2\log\mu$, so Proposition 2 gives the singularity of $d_E(x)$.

We look next how does $d_E(x, \partial \omega)$ 
depends on $E$ in the large.
For fixed $x_0$ away from $\omega$, let 
$\Upsilon_E$ as above, be a minimal $d_E$-geodesic between $N=\partial \omega$
and $x_0$, parametrized with arc-lenght. We can arrange so that $N$
is a level surface for $d_E$. Using variations of geodesics
as in [HeSj], we prove~:
\smallskip
\noindent {\bf Proposition 3}: For all $\e >0$, 
there is a ($\mu$-independent) neighborhood $\Omega_E$
of $\Upsilon_E([0, 1-\e ])$, a $\mu^2$-neighborhood $I_E$ of $E$, such that~:
(i) $(x, E')\mapsto d_{E'}(x, N)\in C^\infty(\Omega_E\times I_E)$.
(ii) $\Omega_E$ is {\it starshaped}, in the following sense~:
$\forall (x, E')\in  \Omega_E\times I_E, \exists ! \ d_{E'}$-minimal geodesic
joining $N$ to $x$ that stays in $\Omega_E$. 
\medskip
\noindent {\bf 3. The tunnel cycle}.
\smallskip
We label objets belonging the to left (resp. right) well with subscript $L$
(resp. $R$). Extend $\Lambda_L=\bigl(\widetilde \Lambda_\iota^E\bigr)_L$ 
along the bicharacteristic flow of $q$. 

For a general, non integrable system, there is no reason for this extension 
coincides with $\Lambda_R=\bigl(\widetilde \Lambda_\iota^E\bigr)_R$. 
However, we say that the pair $(\rho_L, \rho_R)\in \Lambda_L\times
\Lambda_R$,  are 
{\it in correspondance} along a bicharacteristic $\gamma$ if $\Lambda_L$
(or equivalently, because of symmetry, $\Lambda_R$) 
supports a quasi-mode, and
$(\rho_L, \rho_R) \in \gamma\times \gamma$. 
We call the bicharacteristic $\gamma$ a {\it tunnel cycle} 
if there is a pair $(\rho_L, \rho_R)$ in correspondance along $\gamma$, with
$(\rho_L, \rho_R)\in \bigl(\Lambda_\partial^E\bigr)_L
\times \bigl(\Lambda_\partial^E\bigr)_R$. 
Then $\rho_L$ and $\rho_R$ are necessarily umbilics, and $\gamma$
a geodesic between $U_L(E)$ and $U_R(E)$.  
A tunnel cycle will be called {\it minimal} if the geodesic $\gamma$
is minimal, hence of lenght $d_E\bigl(U_L(E), U_R(E)\bigr)$. 
Generically, bicharacteristics connecting
pairs in correspondance and tunnel cycles are discrete sets.
Moreover, pairs in correspondance, in case of Birkhoff tori,
are only defined modulo ${\cal O}(E^\infty)$, since this is the case
for $\widetilde \Lambda_\iota^E$ and $\Lambda_\iota^E$.
Tunnel cycles are exceptional, but as we shall see, there are many pairs in 
correspondance (belonging to different bicharacteristics)
close to the umbilics. 
See [Gr] and [DoSh] for related notions.

\medskip
a) The case of a minimal tunnel cycle.
\smallskip
The picture is the following :
\vskip 2truecm

\includegraphics[height=3.5cm]{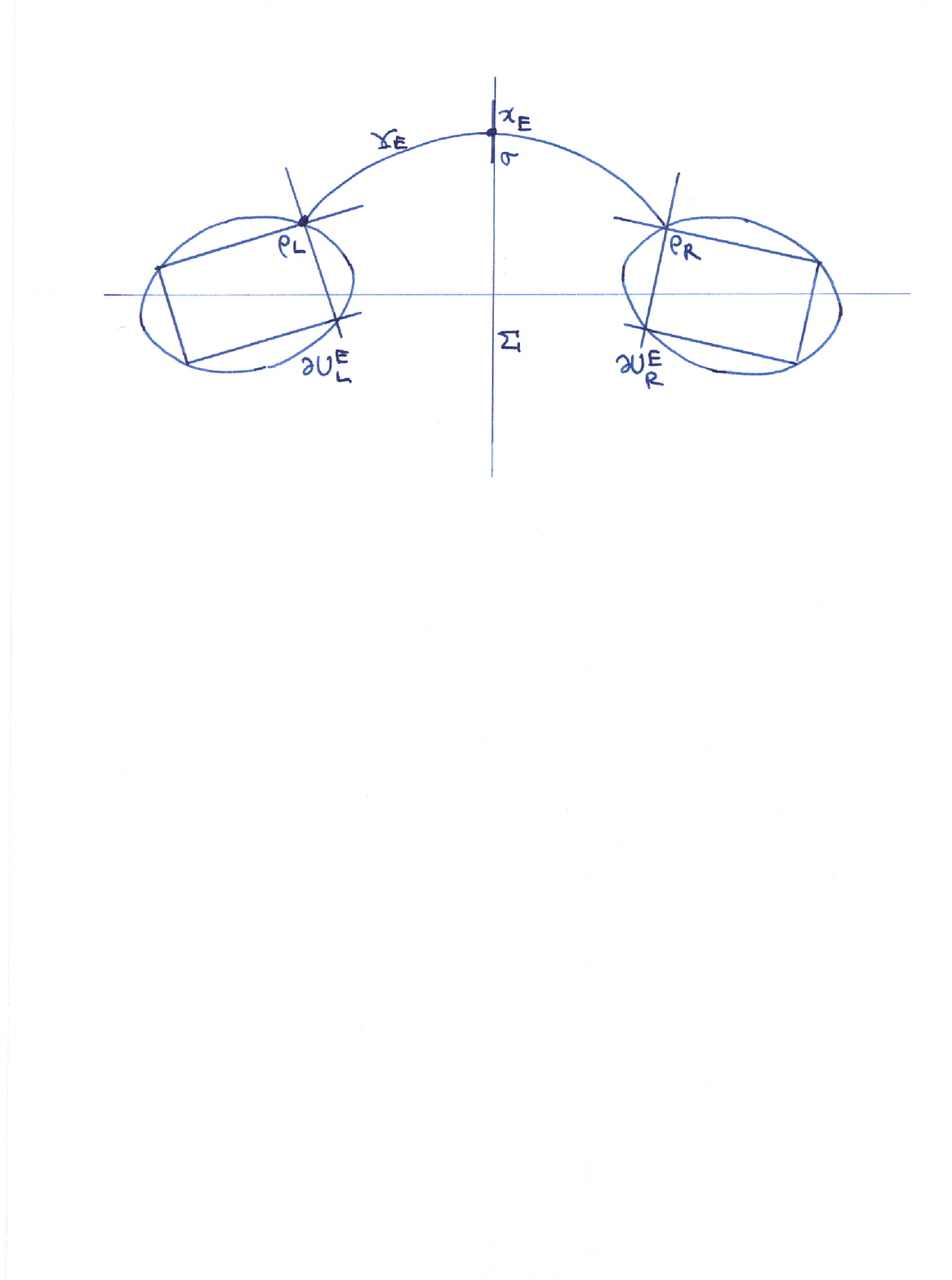}

\vskip 1 truecm

Let $u_L(x, E, h)$ and $u_R(x, E, h)$ be the quasi-modes associated with
the umbilics $\rho_L$ and $\rho_R$, continued beyond the symmetry axis 
$\Sigma$ separating the 2 wells, in a neighborhood of the minimal
geodesic $\Upsilon_E$.
Using Agmon estimates as in [HeSj], we can show that they approximate 
suitably the true eigenfunctions (provided a gap condition.~)
Assume for simplicity there is just
one such minimal geodesic intersecting $\Sigma$ at $x_E$. 
Since $u_L$ and $u_R$ are real near
$\Sigma$, 
the eigenvalue splitting is given by the usual formula
$$
E^+ - E^- = 4h^2 \int_\sigma u_L(0, x_2) {\partial u_R \over \partial x_1}(0,
x_2)dx_2 + {\cal O}(e^{-(S_0 + \e _0)/h}) \leqno(3)
$$
where $\sigma\subset \Sigma$ is a neighborhood of $x_E$.
Denote by $S_L-S_R^*$ the phase that comes up in (3), where $S_L$
and $S_R^*$ stand for suitable $F_y^E(x)$ as above.
By the remark following Proposition 1, $S_L-S_R^*$ has a non degenerate
critical point
precisely at $x_E=\Upsilon_E\cap \sigma$. 
Moreover, the asymptotics of 
the quasi-modes near $U_L(E)$ given in [KaRo], propagate all along
$\Upsilon_E$, 
so the integral can be computed by
standard stationary phase expansion around $x=x_E$. Since the amplitude
of $u_R$ (and $u_L$) is non vanishing, $E^+ - E^-$ is exactly of the
order $e^{-S_0(E)/h}$.

\medskip
b) The general case.
\smallskip

Given $(\rho_L, \rho_R)$ in correspondance, we want to compare,
for $x\in\Sigma$, the
action $F_y^E(x)$ along the bicharacteristic connecting 
$\rho_L$ and $\rho_R$ with Agmon distance $d_{E'}(x)$ relative to
a nearby energy value $E'$. 
For this, let $\Upsilon_{E'}$ be a minimal geodesic between
$U_L(E')$ and $U_R(E')$, intersecting $\Sigma$ at $x_{E'}$, the left and
right components of $\partial U_{E'}$ at $y'_L=y_L(E')$ and $y'_R=y_R(E')$
respectively, and 
consider the {\it lattice of umbilics} carrying quasi-modes around such a  
point. At first approximation, umbilics are of the form
$y=(\lambda_1^{-1}\sqrt {2\lambda_1\iota_1}, 
\lambda_2^{-1}\sqrt {2\lambda_2\iota_2})$, or by (1),
$y=(\lambda_1^{-1}\sqrt {2h\lambda_1\alpha_1}, 
\lambda_2^{-1}\sqrt {2h\lambda_2\alpha_2})$, so the typical neighboring
distance between umbilics is 
$ h ((\alpha_1h)^{-1}+(\alpha_2h)^{-1})^{1/2}$, which is greater than
$h/\mu$, but of the same order when tori are not ``too flat''. 
\vskip 2 truecm
\centerline{\includegraphics[height=5.5cm]{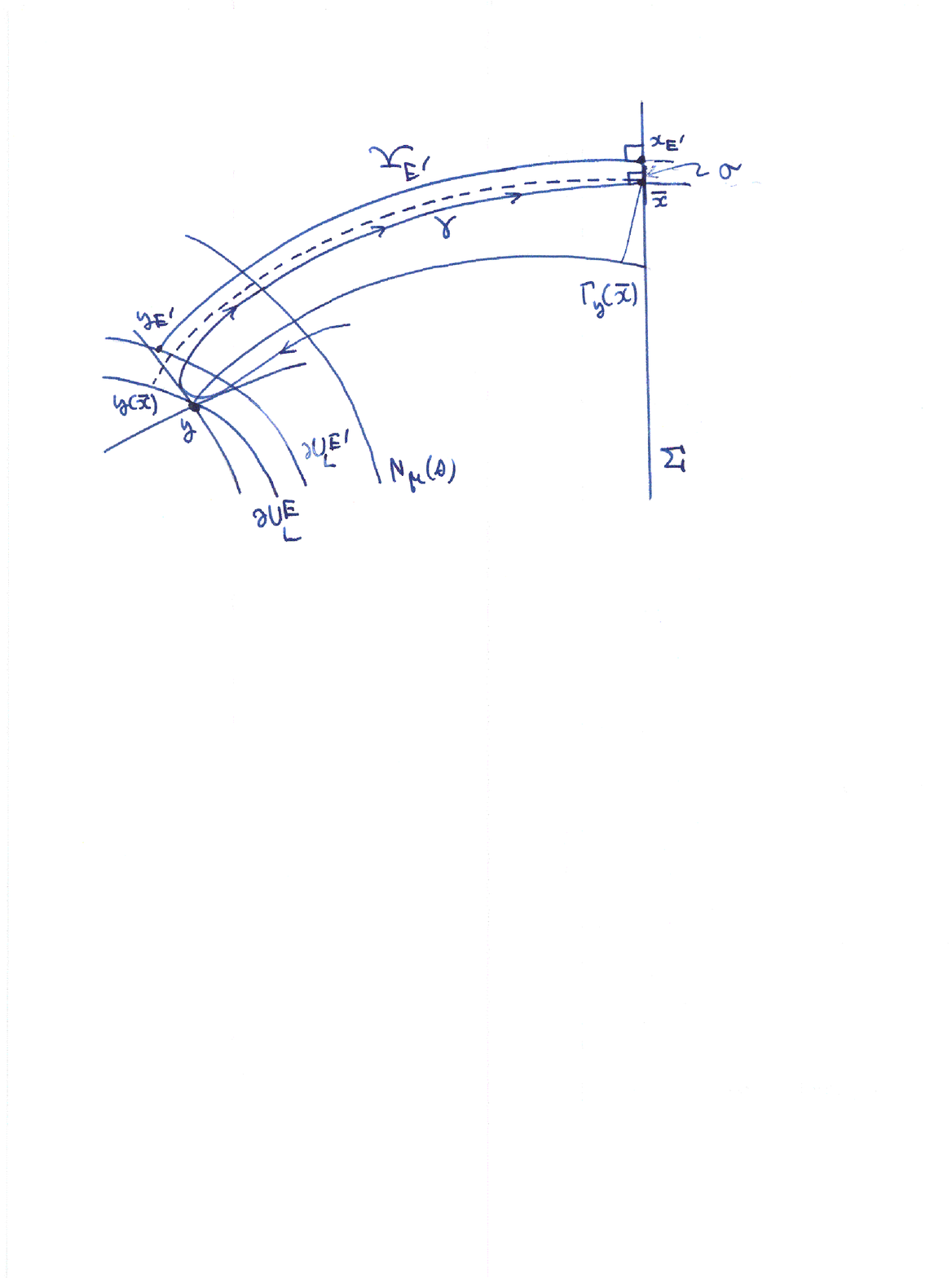}}
\vskip 2 truecm
Let $y$ be such an umbilic, and $\Lambda_L=\bigl(\widetilde 
\Lambda_\iota^E\bigr)_R$ the corresponding Lagrangian manifold.
It is easy to see that 
there is a bicharacteristic $\gamma\subset \Lambda_L\cap\Lambda_R$
(and points in correspondance) such that $S_L-S_R^*$ has a 
non degenerate critical
point $\overline x\in \gamma\cap\Sigma$. 
We have~:
$$(S_L-S_R^*)-S_0(E')=2(F_y^E(\overline x)-d_E(\overline x))
+2(d_E(\overline x)-d_{E'}(\overline x))
+2(d_{E'}(\overline x)-d_{E'}(x_{E'}))$$
and combining Propositions 1-3 gives, under the above hypotheses, that
$(S_L-S_R^*)-S_0(E')=o(1)$, as $h\to 0$, either in case of Birkhoff 
or KAM tori. To compute (3), one has also to know something about
the amplitude, so we need to improve somewhat the expansions of
[KaRo] when getting closer to the caustics.

\medskip
\centerline {\bf References}

\noindent [Cr] S.C. Creagh. Tunneling in two dimensions. Proc. on ``Tunneling
in Complex systems'' (INT 97-1) Seattle, April 30-May 30, 1997.

\noindent [DoSh] S. Dobrokhotov, A. Shafarevich. Math. Phys. Anal. Geom. (2) 
p.141-177, 1999.

\noindent [Gr] A. Grigis. S\'eminaire EDP, Expos\'e XXIII. Ecole Polytechnique.
1994-95.

\noindent [HeSj] B. Helffer, J. Sj\"ostrand. 
Comm. Part. Diff. Eq. 9(4) p.337-408, 1984.

\noindent [KaRo] N. Kaidi, M. Rouleux. 
Comm. Part. Diff. Eq., 27(9 and 10), p.1695-1750, 2002.

\noindent [Ma] A. Martinez. Bull. Soc. Math. France 116 (2), p.199-219, 1988.

\noindent [Po] G. Popov. 
Ann. H. Poincar\'e, Phys. Th. 1(2), p.223-248, 2000. 
Ann. H. Poincar\'e, Phys. Th. 1(2), p.249-279, 2000.

\noindent [Sj] J. Sj\"ostrand.  Asymptotic Analysis, 6, p.29-43, 1992.

\noindent [Wi] M. Wilkinson. Physica 21D, p.341-354, 1986.

\end{document}

%% file: Definitions.tex
\font\twelvec=msbm10 at 12pt
\font\sevenc=msbm10 at 9pt
\font\fivec=msbm10 at 7pt

\newfam\co 
\textfont\co=\twelvec
\scriptfont\co=\sevenc
\scriptscriptfont\co=\fivec

\def\e{\mathop{\rm \varepsilon}\nolimits}

\def\dist{\mathop{\rm dist}\nolimits}

\def\Sum{\displaystyle\sum}

